\documentclass[aps,prl,twocolumn, superscriptaddress, amsmath]{revtex4-1}
\usepackage{natbib}
\usepackage{graphicx}
\usepackage{bm}
\usepackage{color,soul}
\usepackage{hyperref}
\usepackage{float}
\usepackage{csquotes}
\usepackage{color,soul}
\usepackage{hyperref}
\hypersetup{colorlinks=true, urlcolor=blue, citecolor=blue}
\usepackage{longtable}
\usepackage{tabularx}
\usepackage{adjustbox}
\usepackage{multirow} 

\setcitestyle{journalcolor= blue}

\usepackage{babel}

\newcommand{\mpb}{{\rm MHyPbBr$_3$}}
\newcommand{\mpc}{{\rm MHyPbCl$_3$}}

\newcommand{\pb}{{\rm \textit{Pb2$_1$m}}}
\newcommand{\p}{{\rm \textit{P2$_1$}}}
\newcommand{\pmma}{{\rm \textit{Pmma}}}

\begin{document}

\title{Giant Spin Splitting and its Origin in Methylhydrazinium Lead Halide Perovskites}
\author{Nikhilesh Maity}
\email{nikhileshm@usf.edu}
\author{Ravi Kashikar}
\author{S. Lisenkov}
\author{I. Ponomareva}
\email{iponomar@usf.edu}
\affiliation{Department of Physics, University of South Florida, Tampa, Florida 33620, USA}

\date{\today}

\begin{abstract}

Spin splitting, or removal of spin degeneracy in the electronic energy band/level is often a measure of spin-orbit coupling strength and a way to manipulate spin degrees of freedom. We use first-principles simulations to predict giant spin splitting in  methylhydrazinium lead halide (MHyPbX$_3$, MHy = CH$_3$NH$_2$NH$_2$, X = Br and Cl) hybrid organic-inorganic perovskites.  The values can reach up to 408.0~meV at zero Kelvin and 281.6~meV at room temperature. The origin of the effect is traced to the large distortion of PbX$_3$ framework, driven primarily by Pb ions in the ferroelectric $\Gamma^{3-}$ mode. The Pb displacements consist of combination of polar and antipolar arrangements and result in up to 39.2~meV/atom enhancement of the spin-orbit coupling energy in the polar phase of the materials. The spin-orbit coupling gives origin to highly persistent spin textures in MHyPbX$_3$, which are desirable for applications in spintronics and quantum computing. Our findings reveal an additional functionality for hybrid organic-inorganic perovskite and open a way for the design of more materials with giant spin splitting.

\end{abstract}
\maketitle

\section{Introduction}

Spin splitting (SS) is defined as removal of spin degeneracy in the energy level of an isolated system or an energy band of a solid. It often originates from the relativistic effect of spin-orbit coupling (SOC) and was originally investigated by Dresselhaus~\cite{PhysRev.95.568} and Rashba et al.~\cite{1571698600346713472}. As a result, the effect carries their names. The SOC contributes a weak perturbation to the Hamiltonian of the system, which is proportional to $(\boldsymbol{\nabla} V \times  \mathbf{p} )\cdot \boldsymbol{\sigma} $, where $V$ is the crystal potential, $\mathbf{p}$ is the momentum operator, and $\boldsymbol{\sigma}= (\sigma_x,\sigma_y, \sigma_z)$ are the Pauli matrices. Consequently, the effect originates from the presence of (local) electric field, which could be due to the lack of inversion symmetry in bulk material \cite{PhysRev.100.580}, interfacial asymmetry in heterostructures~\cite{Sheka-1959}, and even local asymmetry in centrosymmetric compounds~\cite{Zhang2014}. The effect leads to spin-momentum locking and intriguing physical manifestations, which include 
 intrinsic spin-Hall effect~\cite{Manchon2015}, gate-controlled spin precession~\cite{Supriyo-1990}, (inverse) spin galvanic effects, photogalvanic effects~\cite{Wang2021}, and chiroptic effects~\cite{D0NR05232A}. These are believed to be promising for 
emerging spin-orbitronic devices, that rely on manipulation of the spin degrees of freedom by electrical, optical, or magnetic means~\cite{Manchon2015,Kepenekian2015}.  SS can also serve as a measure of SOC strength, which gives origin to many exotic phenomena, such as the switchable tunneling anomalous Hall effect (TAHE)~\cite{RevModPhys.82.1539,Vedyayev4815866} with ferroelectric polarization and tunneling anisotropic magnetoresistance (TAMR)~\cite{PhysRevLett.98.046601,PhysRevLett.99.056601} with magnetization.

\begin{figure}
\centering
\includegraphics[width=0.5\textwidth]{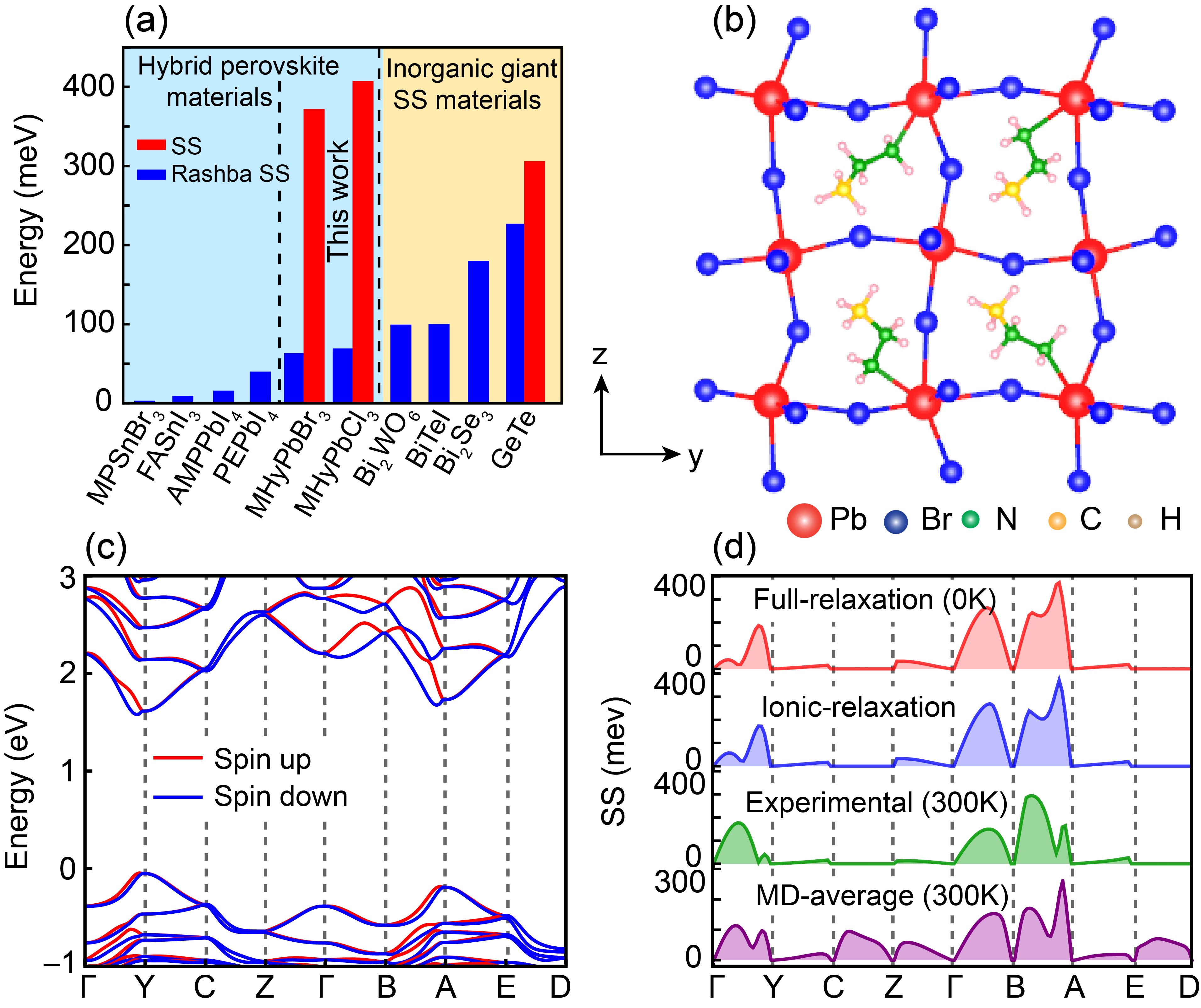}
\caption{(a) (Rashba) SS for some representative materials from literature~\cite{kashikar2022rashba,stroppa2014tunable,Wang2020,Yaxin2017,Djani2019,Ishizaka2011,PhysRevLett.107.096802,Sante2013} and our data. (b) crystal structure and (c) electronic band structure for \p\ phase of \mpb. (d) SS along different directions in the momentum space for \mpb\ structures as obtained from different relaxation schemes of \p\ phase.}
\label{fig1}
\end{figure}

Naturally, these intriguing effects and their emergent and potential applications rely on strong spin-momentum coupling and large SS, which in turn originate from large SOC term \cite{Manchon2015,Lesne2016-mf,Ishizaka2011}. For example, for room temperature applications it is desirable to have a material that exhibits SS in excess of 30~meV. Unfortunately, being weak relativistic effect SOC typically is not strong enough and so far only a handful of materials exhibited experimentally measurable or technologically significant values. One example is polar semiconductor BiTeI, which possesses Rashba SS up to 100 meV \cite{Ishizaka2011}. Rashba SS is defined as the difference between the energy value at the momentum offset and time reversal invariant momentum point~\cite{PhysRevB.102.144106}. The large value in BiTeI is attributed to the loss of inversion symmetry. Another example is cleaved topological insulator Bi$_2$Se$_3$, which exhibits Rashba SS of up to 180~meV, which originates from band-bending induced potential gradient \cite{PhysRevLett.107.096802}. In ferroelectric GeTe, Rashba SS of 227~meV has been predicted computationally~\cite{Sante2013}. This huge SS in GeTe is due to the unusual ordering of the bands near the Fermi surface that, combined with the small band gap, amplifies the effects of spin-orbit interaction~\cite{Sante2013}. Figure~\ref{fig1}(a) compiles Rashba SS values from the literature. 

It should be mentioned that SS can occur also in the absence of SOC in some antiferromagnetic materials. For example, in antiferromagnetic materials lacking  $\theta$$I$ symmetry ($\theta$ represents time reversal and $I$ represents spatial inversion symmetry) and belonging  to the magnetic space group of type III SS can exist without SOC~\cite{PhysRevB.102.014422}. Likewise, altermagnets can exhibit spin splitted bands in absence of SOC owing to a special arrangement of spins and crystal symmetries~\cite{PhysRevX.12.011028}.

Recently, the hybrid organic-inorganic perovskite with chemical formula ABX$_3$, having organic molecule on the A site and, in some cases, organic ligands on the X site are in the focus of scientific attention~\cite{Stroppabook}. Along with their electronic and optoelectronic device applications~\cite{Stranks2015,Park2016,kashikar2022rashba}, their low yield strain~\cite{TU20212765} offers an opportunity to realize emerging devices, such as foldable, bendable, stretchable, and wearable devices~\cite{Kim2010}. For example, MAPbI$_3$ (where MA= CH$_3$NH$_3$) was realised as a flexible nonvolatile memory device~\cite{Gu2016} and flexible diffusive memristor for artificial nociceptors~\cite{Patil2023}. However, these materials typically exhibit rather mediocre SS (see Figure~\ref{fig1}(a)). Among the largest values reported so far is Rashba SS of 40~meV, which was reported for two dimensional hybrid organic-inorganic perovskites, (PE)$_2$PbI$_4$ (where PE= C$_6$H$_5$C$_2$H$_4$NH$_3$)~\cite{Yaxin2017} . Recently, the new hybrid organic-inorganic perovskites MHyPbX$_3$, where MHy is methylhydrazinium molecule CH$_3$NH$_2$NH$_2$ and $X$ is Br or Cl, have been experimentally synthesized~\cite{mpc,mpb}. The \mpb\ undergoes a phase transition from nonpolar \textit{Pm$\bar{3}$m} phase to polar \p\ phase at 418~K, while \mpc\ undergoes polar \pb\ to polar \p\ phase transition at 342~K. Both materials are known to exhibit strong octahedral distortions due to presence of large-sized MHy molecule (ionic radius 2.64~\AA) in comparison with other hybrid perovskites~\cite{mpc,mpb}. Along with the large distortions, the presence of heavy element (Pb) on the B site suggests that this material could exhibit large or even giant SS. Moreover, this material is also ferroelectric which could offer the advantage of Rashba ferroelectricity co-functionality~\cite{kashikar2022rashba}. Having the band gap in the visible region (for example 2.58~eV for \mpb~\cite{mpb}), these materials are also promising for optoelectronics applications. In this work we use first-principles DFT calculations with the aims (i) to predict giant SS of \mpb\ and \mpc; (ii) to reveal its fundamental origin; (iii) to report the unique spin textures in both \mpb\ and \mpc, which are not present in inorganic materials with giant SS. 

\section{Methods}

We used first-principles density functional theory (DFT) based calculations as implemented in Vienna ab initio simulation package (VASP)~\cite{kresse1996efficient,kresse1996efficiency}. The projector augmented wave (PAW) pseudopotential~\cite{blochl1994projector,kresse1999ultrasoft} and Perdew-Burke-Ernzerhof (PBE)~\cite{perdew1996generalized} with D3 dispersion corrections as proposed by Grimme et al.~\cite{grimme2006semiempirical} were used for the calculations. The energy cutoff of plane wave basis was set to 600~eV. For integration inside the Brillouin zone, we chose $\Gamma$-centered Monkhorst-Pack~\cite{monkhorst1976special} k-point mesh of 8$\times$4$\times$4 for both the compounds. 
The structural relaxations were carried out using conjugate gradient algorithm until the ionic forces were less 5~meVÅ$^{-1}$. All the electronic structure calculations were performed including SOC. For spin texture calculations, we used a 21$\times$21 k-point mesh for a given plane in the reciprocal space. The polarization of the systems was calculated using modern theory of polarization as developed by King-Smith and Vanderbilt~\cite{king1993theory}. To avoid ambiguity due to polarization quantum in the calculation, we constructed a rotodistortion path as proposed in Ref.~\cite{kingsland2021structural}. The ISOTROPY suite was used for the space group determination~\cite{stokes2005findsym}. 

It has been established in the literature \cite{frohna2018inversion} that full structural relaxation using DFT does not always accurately predict structures which are not the ground state, which could lead to erroneous predictions for many properties, including SS and related effects. It has previously been shown that for hybrid organic inorganic perovskites \textit{ab initio} molecular dynamics (AIMD) provides results which match experimental ones most closely~\cite{kashikar2022rashba}. We utilized such AIMD approach within the NVT ensemble, simulated using Nose-Hoover thermostat. The run was initialized using the experimental structures with ionic positions fully relaxed. We simulated 12.5~ps time interval. The last 10~ps of the run were used to compute the thermal average structure. For completeness, we have also computed structure by using full structural relaxation, ionic relaxation only (and keeping lattice parameters equal to those in the experiment), and the experimentally obtained structure provided in Refs.~\cite{mpc, mpb}.

\begin{figure*}[t!]
\centering
\includegraphics[width=0.9\textwidth]{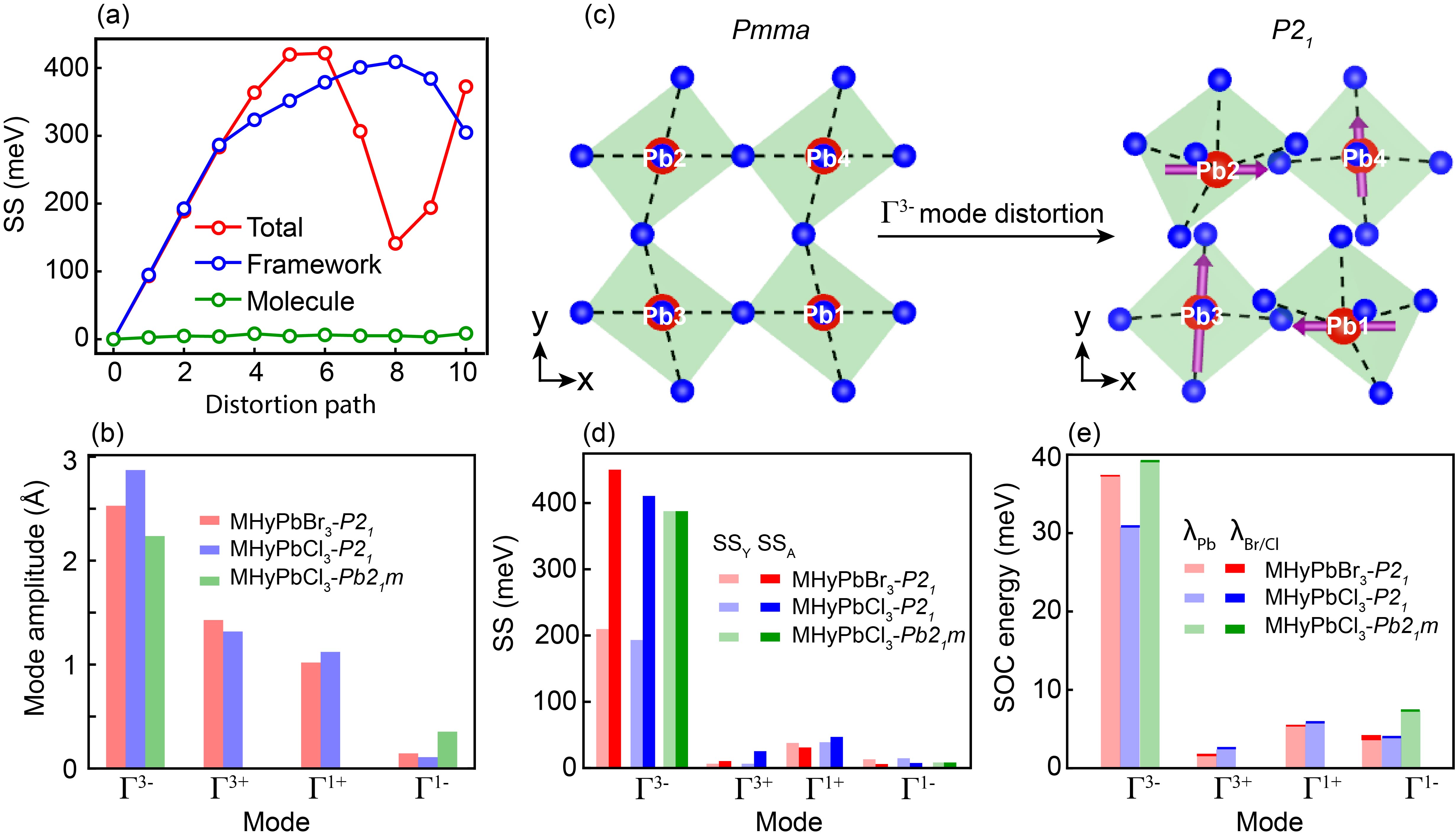}
\caption{(a) Decomposition of the SS at A$^\prime$(0.34, 0.0, 0.5) point in Brillouin zone of \mpb, into contributions from the molecule and framework. (b) Amplitude of structural distortions in different modes of \pb\ and \p\ phases of \mpb\ and \mpc. (c) The atomic displacements in the $\Gamma^{3-}$ mode in \p\ phase of \mpb. (d) The mode decomposition of SS around Y and A high symmetry points (SS$_Y$ and SS$_A$). (e) The effective SOC energies for Pb, Br, and Cl atoms in different distortion modes.}
\label{fig2}
\end{figure*}

\section{Results and Discussion}
We first use DFT calculations to predict the ground state structures of both \mpb\ and \mpc. For that, the experimentally reported structures from Refs.~\cite{mpc, mpb} are subjected to full structural relaxation as outlined above. Both \mpb\ and \mpc\ retained their experimentally reported space groups of \p. A comparison between computational and experimental structural parameters is given in Table S1 of Supplementary Material (SM). The structural files are provided in Ref.~\cite{ourgithub}. The structural visualization of the \p\ phase of \mpb\ and \mpc\ are given in Figure~\ref{fig1}(b) and S1(a), respectively. For \mpc, we also investigate high-temperature polar phase \pb\ (visualized in Figure S1(b)). The spontaneous polarization of the  \mpb\ ground state is 5.66~$\mu$C/cm$^2$. For \mpc\ it is 6.44~$\mu$C/cm$^2$ for \p\ phase and 7.96~$\mu$C/cm$^2$ for \pb\ phase (See Figure S6 of SM), which is in agreement with the recent computational results \cite{srivastava2023guestinduced}.

Next, we compute electronic structures of \mpb\ and \mpc\ taking into account SOC. Figure~\ref{fig1}(c) gives the electronic structure of \p\ phase of \mpb, while data for other phases and materials are given in Figure S3. The blue and red colours indicate bands with opposite spins. The band gaps for \mpb\ and \mpc\ in the monoclinic phases are 1.6~eV and 2.1~eV, respectively. In both materials, the valence band maximum (VBM) and conduction band minimum (CBM) occurs around Y(0.5, 0, 0). Furthermore, in \mpb(\mpc), the eigenvalues of lowermost CB and uppermost VB around A(0.5, 0, 0.5) are 95.2(90.0)~meV greater and 138.4(151.7)~meV lesser than that of Y, which allows for indirect transition between these two points. The partial density of states (Figure S4) reveals that the valence bands near the Fermi level are dominated by X-p and Pb-s orbitals. The conduction bands near the Fermi level are formed primarily from  Pb-p orbitals. The band structures reveal the presence of SS in these materials. To quantify such SS we computed the energy difference between the spin splitted bands along different directions in the Brillouin zone and report it in Figure~\ref{fig1}(d) for \mpb\ and in Figure S5(a) for \mpc. The data reveal giant SS, which can reach 372.6 and 408.0~meV for \mpb\ and \mpc, respectively. These values of SS are among the largest (likely the largest) reported in the literature so far to the best of our knowledge.

To eliminate the possibility that the giant SS is an artifact of simulations, we computed SS for the structures obtained from other relaxation techniques described in the method section, which in addition, allows us to assess the temperature evolution of SS. Figure~\ref{fig1}(d) and Figure S5(a) show a comparison of the SS between such simulations and reveal that all of them predict giant values for SS. The AIMD data correspond to 300~K and indicate that the maximum SS decreases only slightly (down to 281.6~meV) at finite temperatures, which is promising for potential applications. The maximum Rashba coefficient is predicted to be 1.78~eV\AA\ in MHyPbX$_3$, which is in the range of strong Rashba coefficient as classified in Ref.~\cite{MERAACOSTA2020145}. 

\begin{figure}[t!]
\centering
\includegraphics[width=0.5\textwidth]{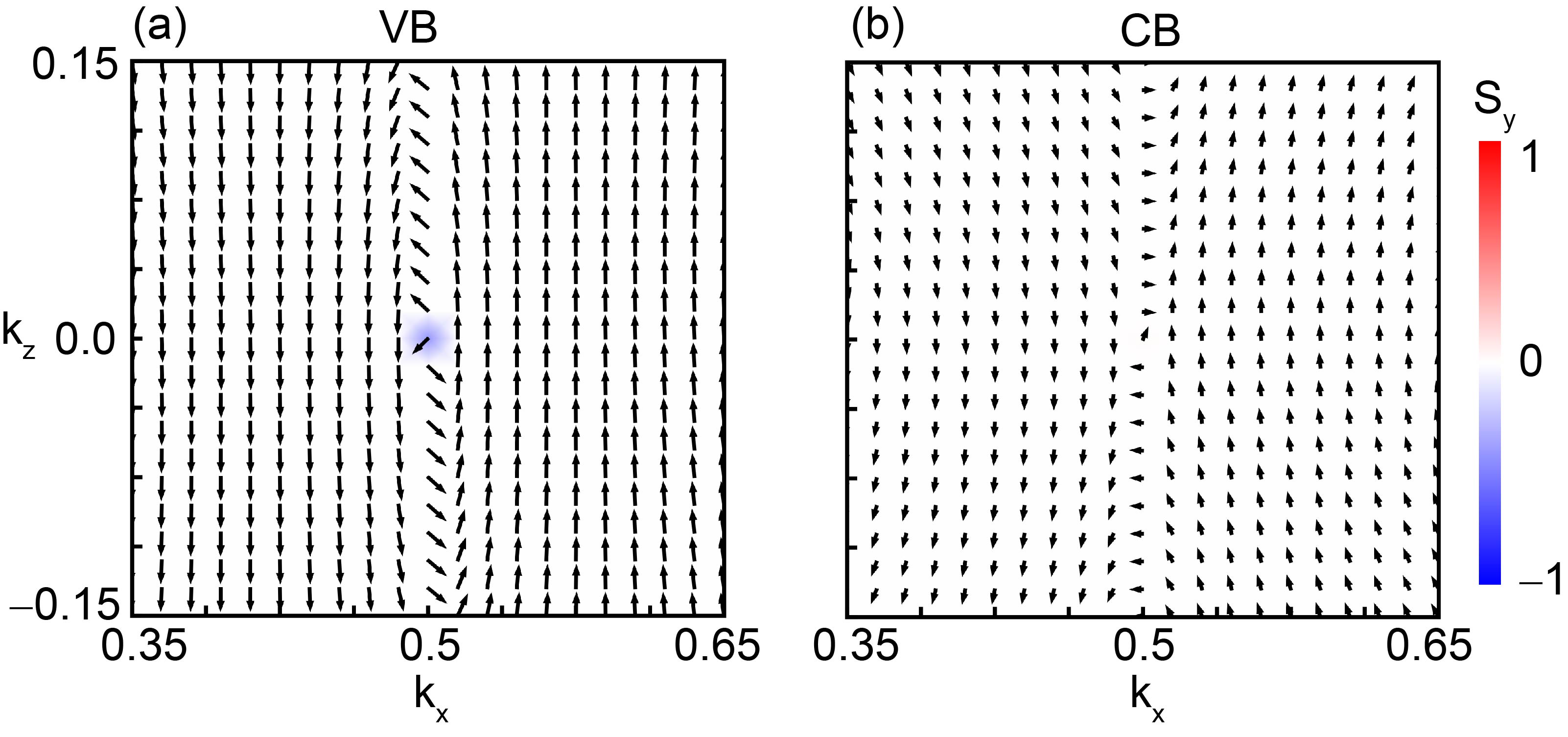}
\caption{The spin texture of (a) VB and (b) CB of \mpb\ around Y high symmetry point in the k$_x$-k$_z$ plane, obtained from DFT. }
\label{fig3}
\end{figure}

To understand the origin of such giant SS in MHyPbX$_3$, we take advantage of computations and decompose SS into the contributions from  MHy molecule and PbX$_3$ framework. For that, we first construct the centrosymmetric phase (\pmma) of the material using PSEUDO~\cite{capillas2011new}.  Next, we connect such centrosymmetric phase to the given noncentrosymmetric phase of the material (\p\ phase for example). A roto-distortion path is generated between the two, which consists of a rotation of the MHy molecule and distortion of PbX$_3$ framework. Such roto-distortion path is then ``decomposed" into two. One of them contains only the rotation of the molecules and keeps the framework ions in their centrosymmetric positions. Another one includes only distortion of the framework and keeps the ions of the molecules in their centrosymmetric positions. These two paths quantify the contributions of the molecule and frameworks, respectively, to the structural distortions associated with the noncentrosymmetric phases~\cite{DiSante2013,Stroppa-Angewandte}. For structures along each of the paths, we compute band structures and associated SS. The same symmetry directions in the Brillouin zone as shown in Figure~\ref{fig1}(c) were considered in each case. Figure~\ref{fig2}(a) reports the evolution of the SS in A$^\prime$(0.34, 0.0, 0.5) point of the Brillouin zone along the distortion path for \mpb. The evolution of SS for \mpc\ is given in Figure S5(b). It reveals that the SS in MHyPbX$_3$ is dominated by the framework. 

\begin{table*}
\begin{center}
\caption{Effective masses ($m_x$, $m_y$, and $m_z$) and spin-splitting strengths ($\alpha_x$, $\alpha_z$, $\beta$, $\gamma_x$, and $\gamma_z$) in unit of free electron mass and meV\AA, respectively.}
\begin{tabular}{ p{2cm} p{1.5cm} p{1.5cm}  p{1.5cm}  p{1.5cm} p{1.5cm} p{1.5cm} p{1.5cm} p{1.5cm} p{1.5cm}}
\hline \hline
\multirow{1}{*}{Compound} & & \multicolumn{1}{c} {$m_x$} & \multicolumn{1}{c}{$m_y$} & \multicolumn{1}{c}{$m_z$}  & \multicolumn{1}{c}{$\alpha_x$} & \multicolumn{1}{c}{$\alpha_z$} & \multicolumn{1}{c}{$\beta$} & \multicolumn{1}{c}{$\gamma_x$} & \multicolumn{1}{c}{$\gamma_z$} \\
\hline 
MHyPbBr$_3$ & \hspace{0.5cm}CB & \hspace{0.5cm}0.27  & \hspace{0.5cm}0.28 & \hspace{0.5cm}0.91  & \hspace{0.5cm}0.1 & \hspace{0.5cm}1.4  & \hspace{0.5cm}0.02 & \hspace{0.5cm}0.01 & \hspace{0.5cm}0.38\\
            & \hspace{0.5cm}VB & \hspace{0.5cm}0.37  & \hspace{0.5cm}0.32 & \hspace{0.5cm}0.71  & \hspace{0.5cm}0.05 & \hspace{0.5cm}0.45  & \hspace{0.5cm}0.04 & \hspace{0.5cm}0.04 & \hspace{0.5cm}-0.01\\

MHyPbCl$_3$ & \hspace{0.5cm}CB & \hspace{0.5cm}0.36  & \hspace{0.5cm}0.36 & \hspace{0.5cm}1.15  & \hspace{0.5cm}0.1 & \hspace{0.5cm}1.34  & \hspace{0.5cm}0.02 & \hspace{0.5cm}0.06 & \hspace{0.5cm}0.28\\
            & \hspace{0.5cm}VB & \hspace{0.5cm}0.44  & \hspace{0.5cm}0.39 & \hspace{0.5cm}0.72  & \hspace{0.5cm}-0.01 & \hspace{0.5cm}0.22  & \hspace{0.5cm}0.03 & \hspace{0.5cm}-0.04 & \hspace{0.5cm}-0.04\\

\hline \hline
\end{tabular}
\label{T1}
\end{center}
\end{table*}

Next, we aim to find out if there is a particular structural distortion of the framework that gives origin to the giant SS. For that, the structural distortion associated with the framework is decomposed into individual modes using ISODISTORT~\cite{stokes2016isosubgroup,campbell2006isodisplace}. The molecules are kept in their centrosymmetric positions. The framework distortion is made up of four $\Gamma$ modes, i.e., $\Gamma^{1-}$$\oplus$$\Gamma^{1+}$$\oplus$$\Gamma^{3-}$$\oplus$$\Gamma^{3+}$. The amplitude of these modes is given in Figure~\ref{fig2}(b) and reveals that  $\Gamma^{3-}$  mode has the largest amplitude. This is also the only ferroelectric mode in this material. To quantify the contributions of these different modes to SS we compute SS associated with a condensation of each individual mode while keeping the molecules in their centrosymmetric positions. The data are given in Figure~\ref{fig2}(d) and reveal that it is the ferroelectric $\Gamma^{3-}$ mode that is responsible for the giant SS in this family of materials. Figure~\ref{fig2}(c) visualizes ionic displacements associated with $\Gamma^{3-}$ mode. The mode is dominated by Pb displacements, which are arranged in both polar and antipolar patterns. To elucidate the mechanism behind ferroelectric mode-driven giant SS, we compute effective SOC energies, $E_{SOC}$, inside the augmentation spheres centered around a given ion as outlined in Refs.~\cite{steiner2016calculation,lenthe1993relativistic}. The difference in these energies between the given phase and the centrosymmetric phases, $\lambda(i)=E_{SOC}(i)-E_{SOC}^{cs}(i)$, for different types of ions are reported in Figure~\ref{fig2}(e). There exists a drastic enhancement of $E_{SOC}$ for the Pb ions in $\Gamma^{3-}$ mode, which explains why this polar mode is responsible for the giant SS. Thus, our calculations reveal that highly distorted PbX$_3$ framework results in large polar displacement of Pb, which causes strong enhancement of spin-orbit interactions, resulting in giant SS. Since the large framework distortion is believed to be caused by the large size of MHy molecule, we believe that our insight will open a way to design of more hybrid organic-inorganic perovskites with giant SS.

SS is known to give origin to spin textures, which are vectorial representations of expectation values of the spin components in momentum space. In ferroelectrics, they can be controlled by the application of an electric field, thanks to the coupling between spin textures and electric polarization~\cite{PhysRevB.102.144106}. Figure \ref{fig3} shows the spin textures in $k_x$-$k_z$ plane for VB and CB around Y high symmetry point of \mpb, which are highly persistent along \textbf{z} direction in the reciprocal space. The persistent spin textures provide a long spin lifetime of carriers, which is highly desirable for spin–orbitronics~\cite{manchon2015new}. For other plane, the spin textures of \mpb\ and \mpc\ are given in Figures S9, S10 and S11. To elucidate the origin of the highly persistent nature of spin textures in \mpb\ and \mpc\, we develop effective Hamiltonian using the method of invariants~\cite{tao2021perspectives}. The point group corresponding to \p\ space group is $C_2$. The transformation rules for wave vector $\boldsymbol{k}$ and spin $\boldsymbol{\sigma}$ in $C_2$ point group are given in Table S2. From the common invariant terms, the effective Hamiltonian is

\begin{eqnarray}
H=  \frac{(\hbar k_x)^2}{2m_x} + \frac{(\hbar k_y)^2}{2m_y} + \frac{(\hbar k_z)^2}{2m_z} + \alpha_x k_x \sigma_x + \alpha_z k_x \sigma_z \nonumber\\
+ \beta k_y \sigma_y + \gamma_x k_z \sigma_x + \gamma_z k_z \sigma_z
\label{equ1}
\end{eqnarray}

Where $k_i$ are components of wave vector, $m_i$ are effective masses, and $\sigma_i$ are the Pauli spin matrices. $\alpha_x$, $\alpha_z$, $\beta$, $\gamma_x$, and $\gamma_z$ are spin-momentum coupling parameters that can be obtained by fitting the eigenvalues of the Hamiltonian of (Eq.~\ref{equ1}) to the values computed from DFT. The parameters are listed in Table~\ref{T1}. An analysis of the parameters' strength indicates that the effective masses are $\boldsymbol{k}$-dependent and have larger values along the $k_z$ direction. In case of the MHyPbX$_3$, we find that among the parameters that control SS, the largest values are for $\alpha_z$ followed by $\gamma_z$, and the other two terms are significantly smaller. For the VB, $\alpha_z$ is nearly order of magnitude larger than other parameters. In this case the Hamiltonian can be approximated as $ \frac{(\hbar k_x)^2}{2m_x} + \frac{(\hbar k_y)^2}{2m_y} + \frac{(\hbar k_z)^2}{2m_z} + \alpha_z k_x \sigma_z$ with the eigenvalues of $E_{\pm}= \frac{(\hbar k_x)^2}{2m_x} + \frac{(\hbar k_y)^2}{2m_y} + \frac{(\hbar k_z)^2}{2m_z}  \pm \alpha_z k_x$ and the associated  expectation value for the spin ${<\mathbf  \sigma}_{\pm}>=(0,0,\pm 1)$. The independence of the expectation value of the $\boldsymbol{k}$-vector (except for the sign flip due to the change of the outer branch from  $E_+$ to $E_-$ as the direction of $k_x$ flips) explains the persistent nature of the spin textures. The finite values of other parameters alter the spin texture near the CBM and VBM slightly toward a quasi-persistent one. 

\section{Conclusions}
In summary, we used DFT calculations to predict giant SS in  MHyPbX$_3$ hybrid organic-inorganic perovskites. The values up to 408.0~meV were obtained for SS at zero Kelvin and up to  281.6~meV in MHyPbCl$_3$ at room temperature. The origin of such a large value was traced to the Pb displacements in the ferroelectric $\Gamma^{3-}$ mode triggered by large framework distortion required to accommodate large MHy molecule. The displacement can reach 0.92~\AA, which combined with the large $Z$ of Pb leads to the strong enhancement of spin-orbit interaction energy, up to 39.2~meV/atom. Moreover, the SOC leads to the development of highly persistent spin textures in the momentum space, which are known to significantly enhance spin relaxation time, and, therefore, are very valuable for applications in spintronics and quantum computing. Thus, our study reveals that  MHyPbX$_3$ is ferroelectric semiconductor with giant SS and highly persistent spin textures in momentum space. All these are the key properties for emerging applications such as in efficient spin field effect transistors, valley spin valve, and barrier of ferroelectric tunnel junctions. In addition, being hybrid organic-inorganic perovskites these materials are relatively soft and flexible, which could bring the realization of flexible, foldable, and even wearable spintronics. The knowledge on the fundamental origin of the giant SS in MHyPbX$_3$ is likely to promote design of other hybrid materials with giant SS and could open a new direction in hybrid perovskites research. 

\section{Acknowledgments}
This work was supported by the U.S. Department of Energy, Office of Basic Energy Sciences, Division of Materials Sciences and Engineering under Grant No. DE-SC0005245. Computational support was provided by the National Energy Research Scientific Computing Center (NERSC), a U.S. Department of Energy, Office of Science User Facility located at Lawrence Berkeley National Laboratory, operated under Contract No. DE-AC02-05CH11231 using NERSC Award No. BES-ERCAP-0025236.x

%

\newpage
\onecolumngrid
\begin{center}
   \textbf{\Large Supplementary Material}
\end{center}

\renewcommand{\thefigure}{S\arabic{figure}}
\setcounter{figure}{0}

\renewcommand{\thetable}{S\arabic{table}}
\setcounter{table}{0}

\begin{figure}[h!]
\centering
\includegraphics[width=0.9\textwidth]{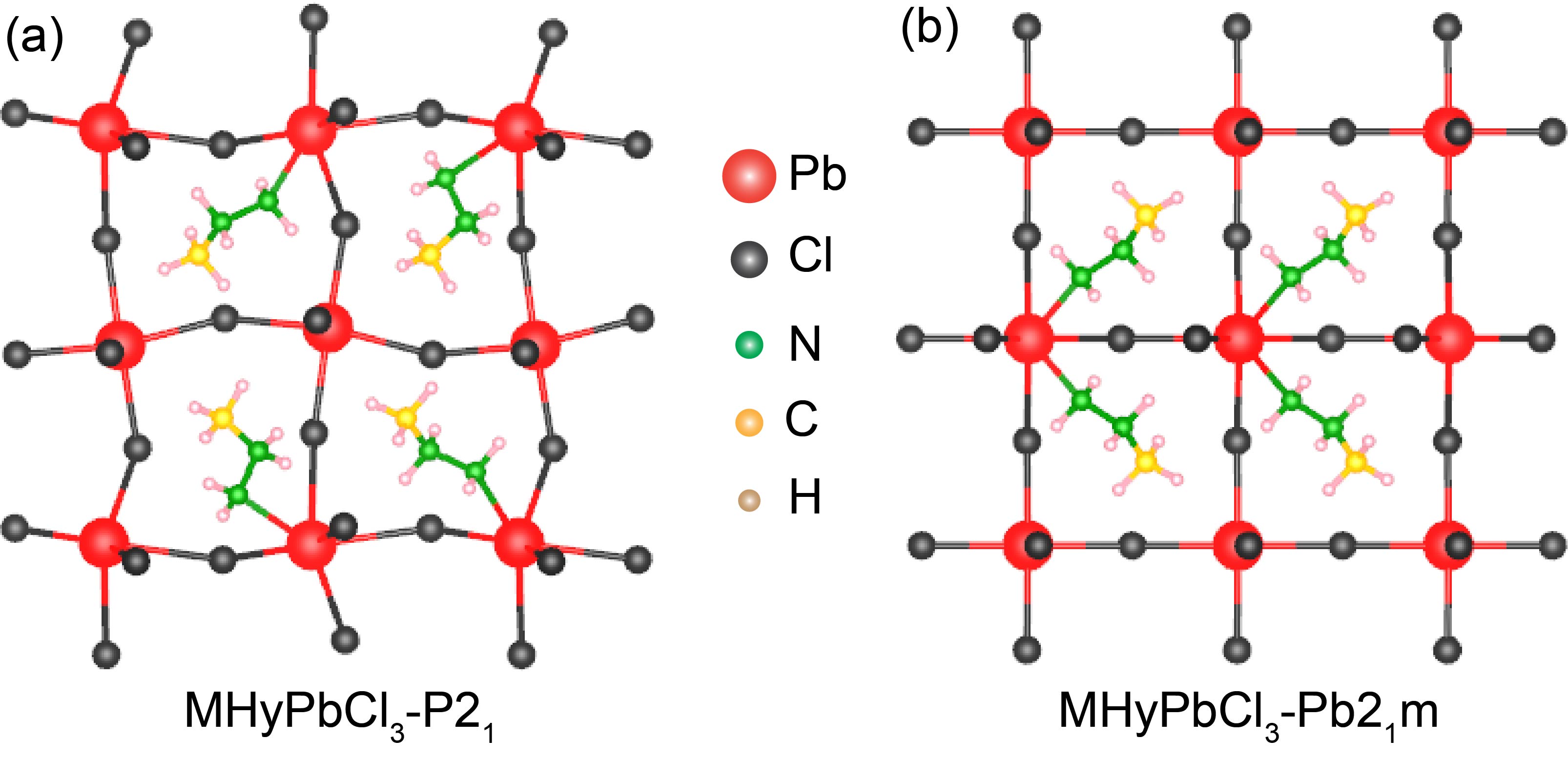}
\caption{(a) Crystal structure for (a) \p\ and (b) \pb\ phase of \mpc.}
\label{Fig1}
\end{figure}

\begin{figure}[h!]
\centering
\includegraphics[width=0.9\textwidth]{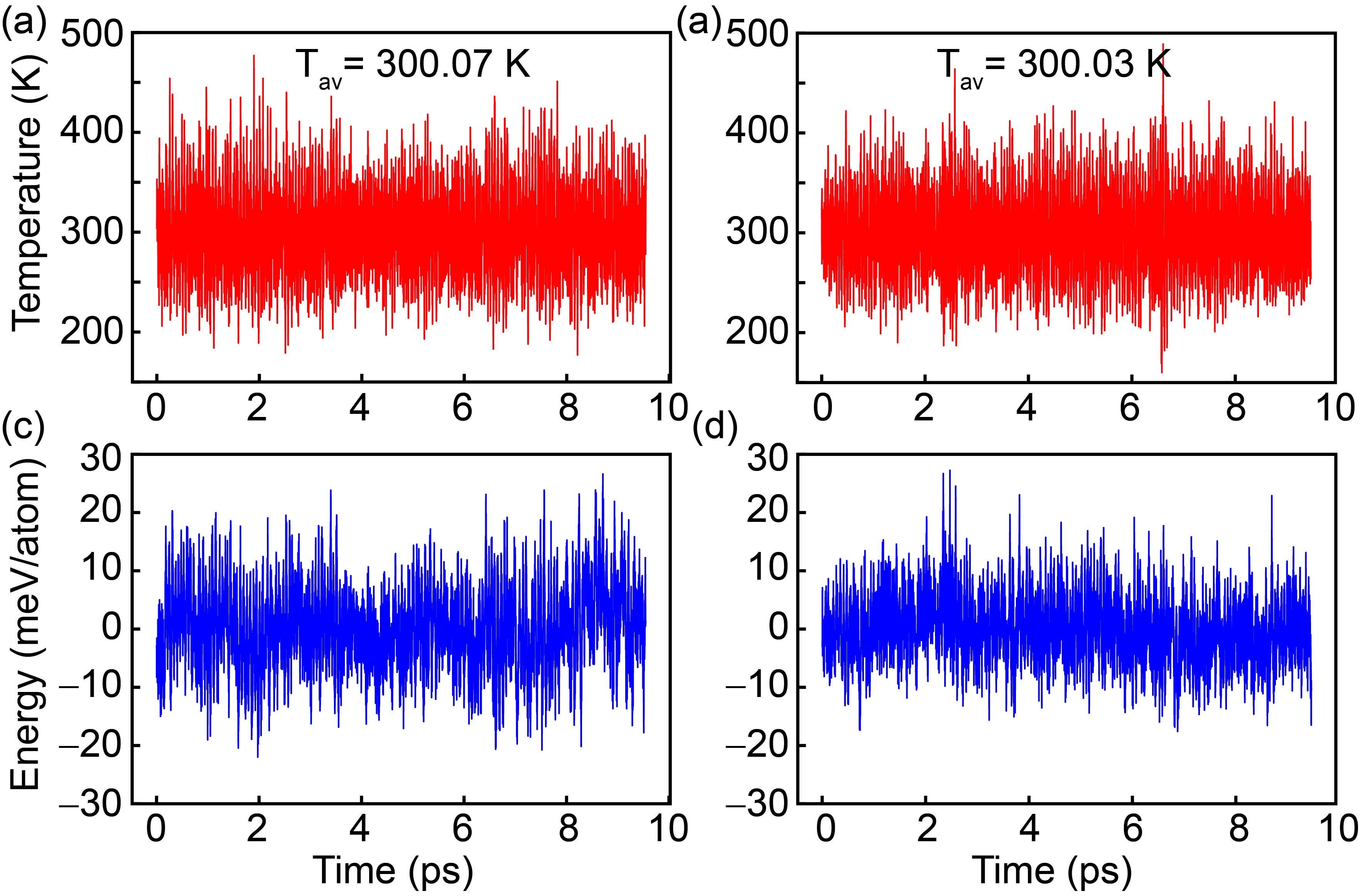}
\caption{Fluctuation of temperature from the equilibrium temperature of 300 K for (a) \mpb\ and (b) \mpc\ for the last 10 ps AIMD steps. Variation of potential energy per atom as a function of the last 10 ps AIMD step at 300K for (c) \mpb\ and (d) \mpc.}
\label{Fig2}
\end{figure}

\begin{longtable}{cc|c|c}
\caption{Structural parameters, total energy, and descriptor comparing various structures. }
\label{tab-s1}\\
\hline
\textbf{Components} && \textbf{MHyPbBr$_3$ ($P2_1$)} & \textbf{MHyPbCl$_3$ ($P2_1$) } \\
\hline
\endfirsthead
\multicolumn{4}{c}%
{\tablename\ \thetable\ -- \textit{Continued from previous page}} \\
\hline
\textbf{Components} && \textbf{Monoclinic ($P$c)} & \textbf{Orthorhombic ($P$na2$_1$) }  \\
\hline
\endhead
\hline \multicolumn{4}{r}{\textit{Continued on next page}} \\
\endfoot
\hline
\endlastfoot
Lattice Parameters(\AA) & Fully relaxed  & 6.00, 11.83, 11.73  & 5.75, 11.33, 11.34 \\
                        & Experimental   & 5.97, 11.83, 11.86  & 5.73, 11.34, 11.39 \\

Monoclinic angle & Fully relaxed       &   91.97 $^\circ$  & 92.26 $^\circ$ \\
                 & Experimental        &   92.36 $^\circ$  & 92.33 $^\circ$ \\

Volume  (\AA$^3$ )   &Fully relaxed   & 832.34 &  738.01 \\

                     & Experimental   & 836.73 &  739.19 \\
\hline

                  & Experimental       & 3.89 & 4.58 \\
                  & Ensemble-Average   & 5.99 & 1.09 \\
Energy (meV/f.u.) & Ionically relaxed  & 1.76 & 0.34 \\
                  & Fully relaxed (Reference)     & 0.00 & 0.00  \\
\hline
                                  & Experimental (Reference)   & 0.00 & 0.00 \\
                                  & Ensemble-Average    & 0.26 & 0.29 \\
Structural descriptor ($\Delta$)  & Ionically relaxed   & 0.18 & 0.09 \\
                                  & Fully relaxed       & 0.20 & 0.10  \\
\end{longtable}

\begin{figure}[h!]
\centering
\includegraphics[width=0.9\textwidth]{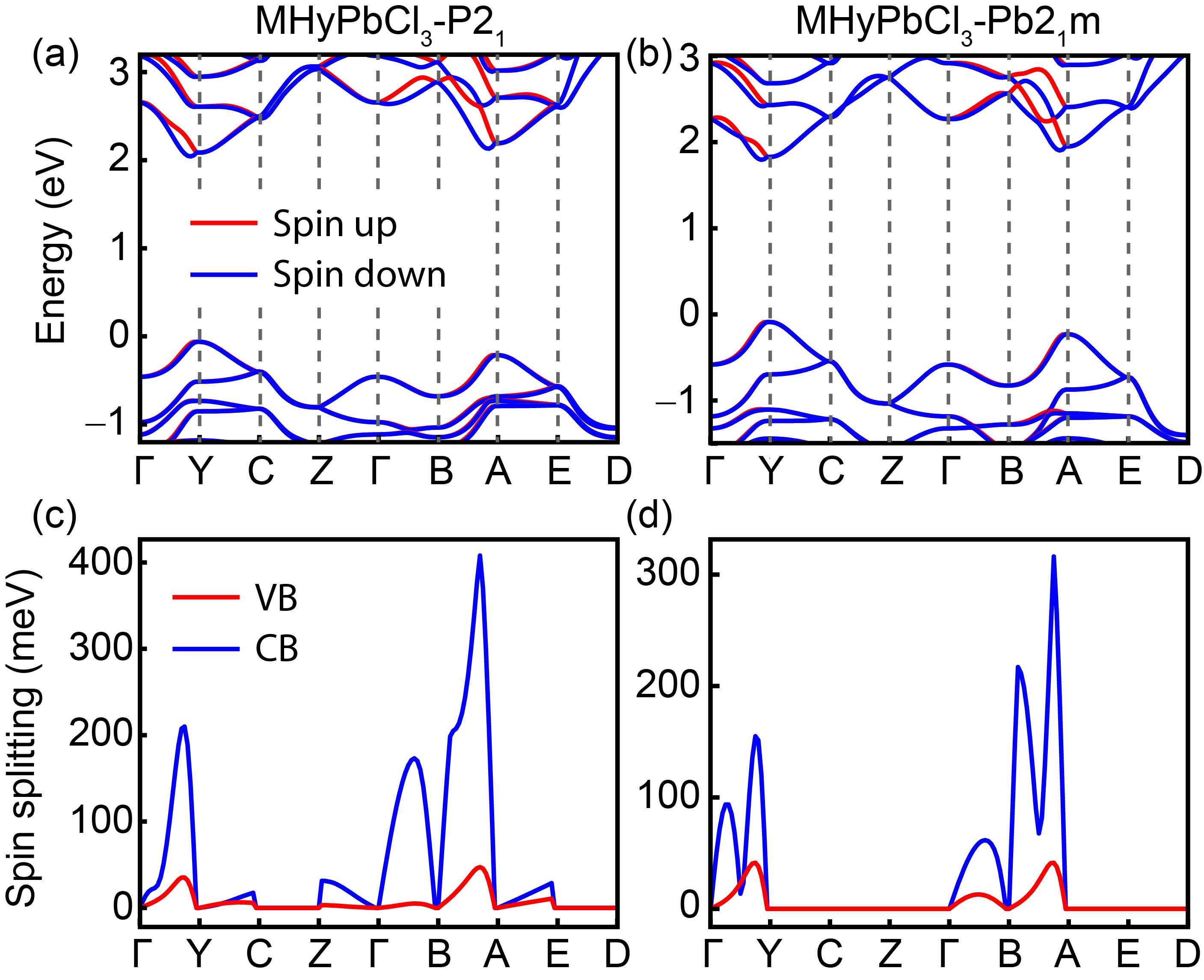}
\caption{Electronic band structure of \mpc\ for (a) \p\ and (b) \pb\ phase. SS of \mpc\ along the symmetry direction in the momentum space for (c) \p\ and (c) \pb\ phase.}
\label{Fig3}
\end{figure}

\begin{figure}[h!]
\centering
\includegraphics[width=0.9\textwidth]{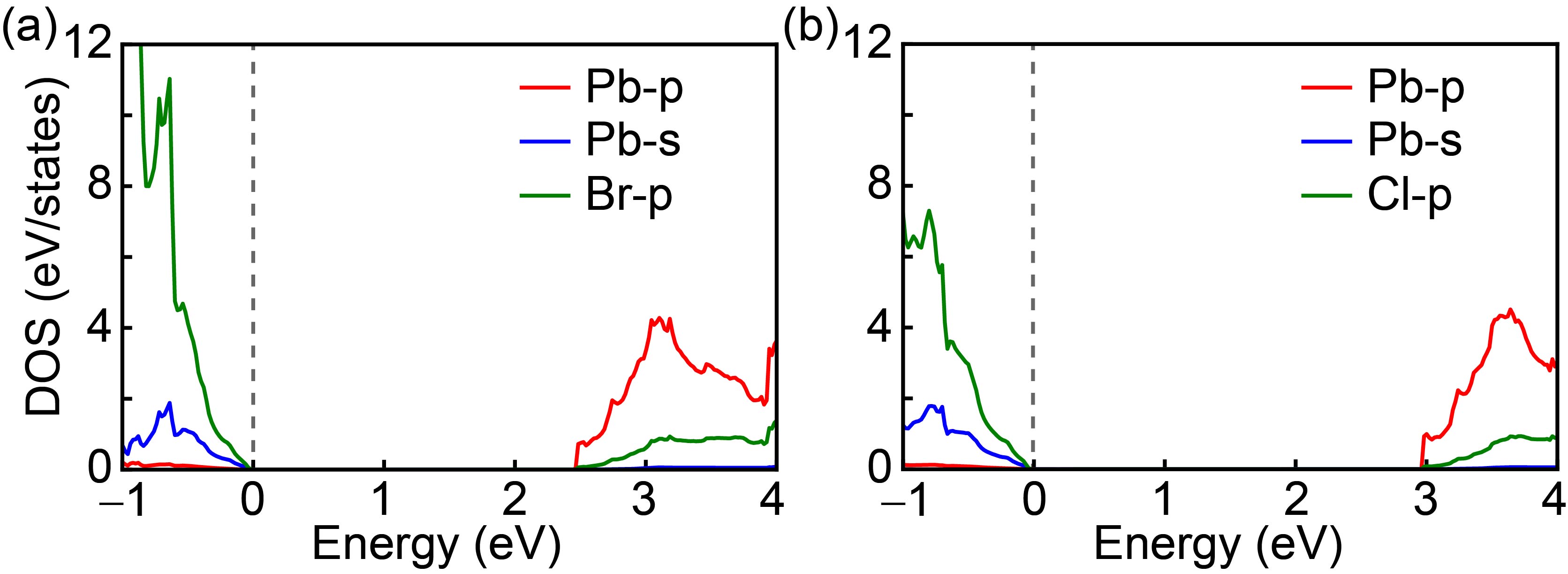}
\caption{Density of states (DOS) for (a) \mpb\ and (b) \mpc.}
\label{Fig4}
\end{figure}

\begin{figure}[h!]
\centering
\includegraphics[width=0.8\textwidth]{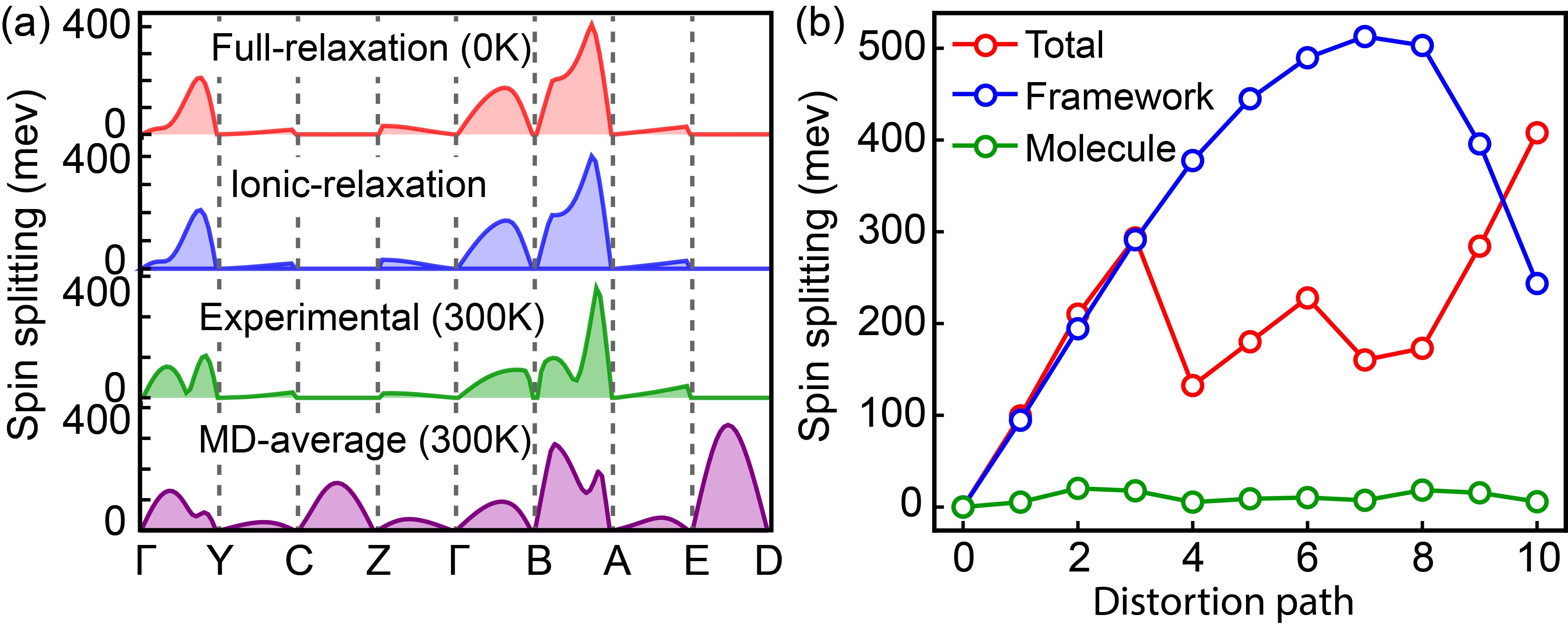}
\caption{(a) SS along different directions in the momentum space for \mpc\ structures as obtained from different relaxation schemes. (b) Decomposition of the SS at A$^\prime$(0.37, 0.0, 0.5) point in Brillouin zone of \mpc, into contributions from the molecule and framework.}
\label{Fig5}
\end{figure}

\begin{figure}[h!]
\centering
\includegraphics[width=1.0\textwidth]{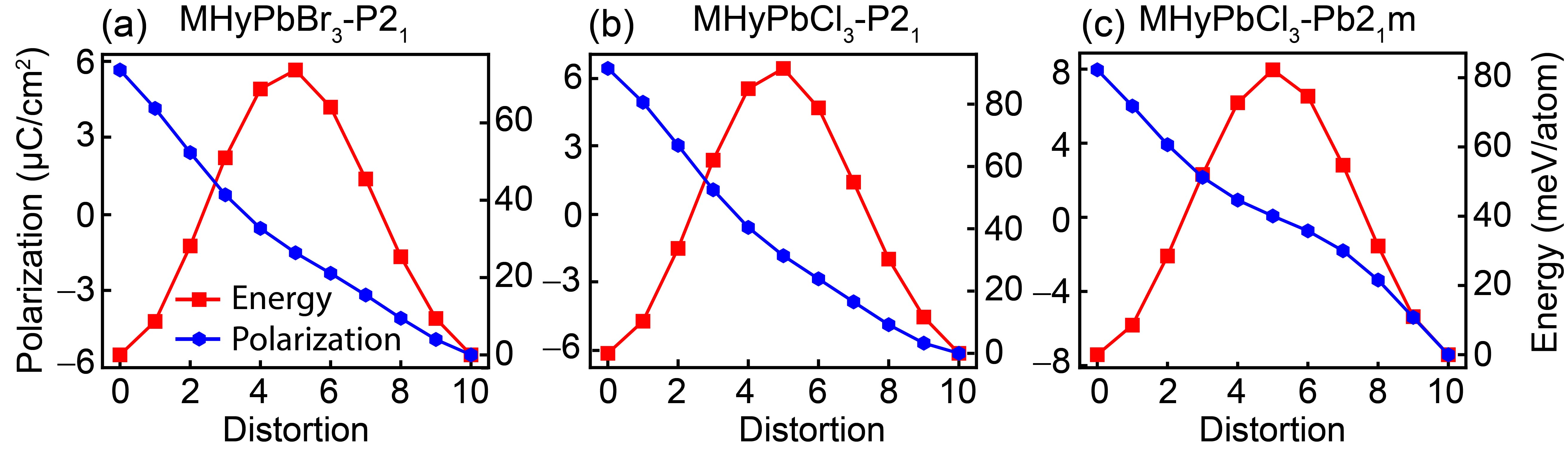}
\caption{(a)-(c)Variation of spontaneous polarization and total energy/atom along roto-distortion path for respective polar phases of \mpb\ and \mpc\ structures.}
\label{Fig6}
\end{figure}

\begin{figure}[h!]
\centering
\includegraphics[width=1.0\textwidth]{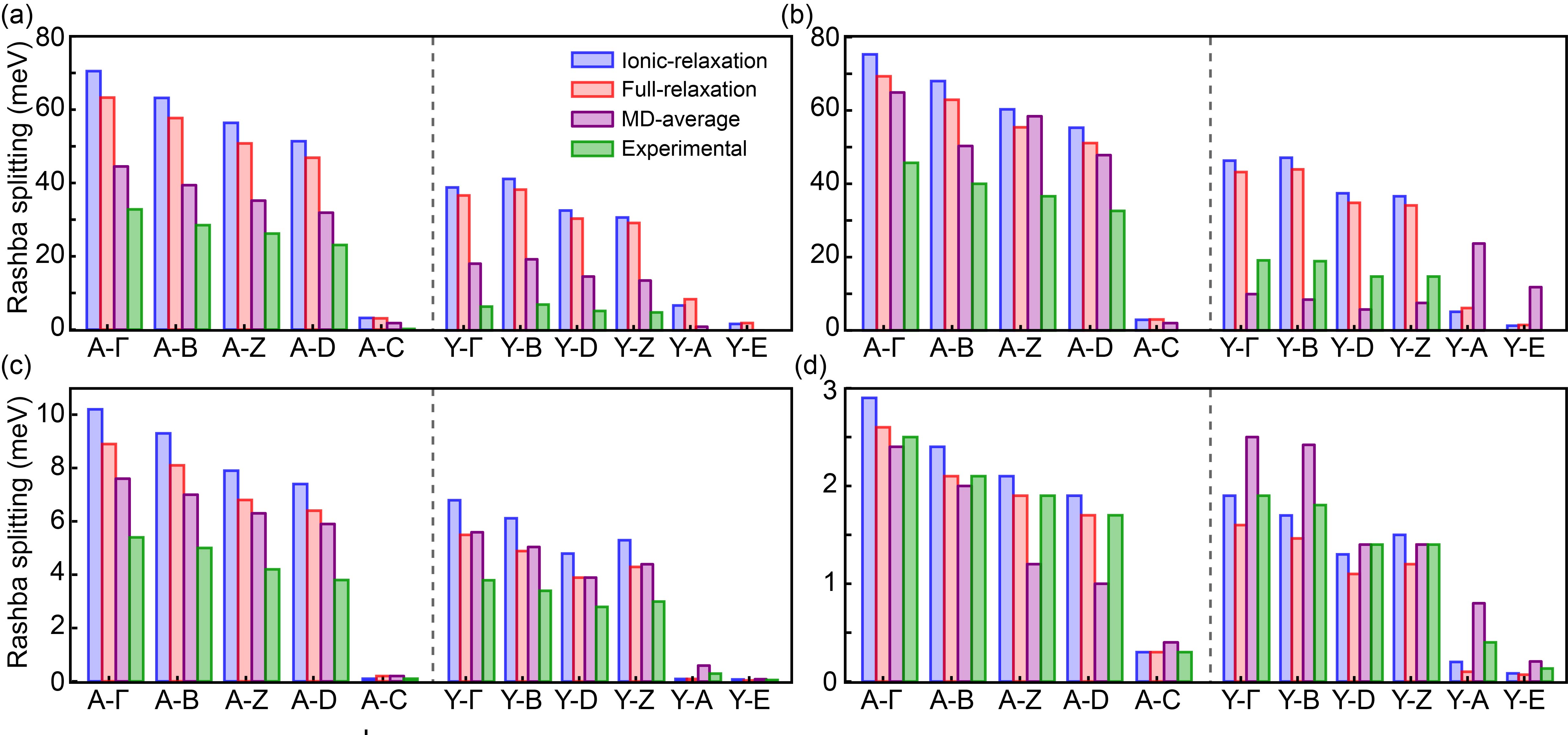}
\caption{Rashba SS along different symmetric direction from the origin of time reversal invariant points (TRIM) A and Y for conduction band (CB) of (a) \mpb\ and (b) \mpc. Rashba SS along different symmetry direction from the origin of time reversal invariant points (TRIM) A and Y for valence band (VB) of (c) \mpb\ and (d) \mpc.}
\label{Fig7}
\end{figure}

\begin{figure}[h!]
\centering
\includegraphics[width=1.0\textwidth]{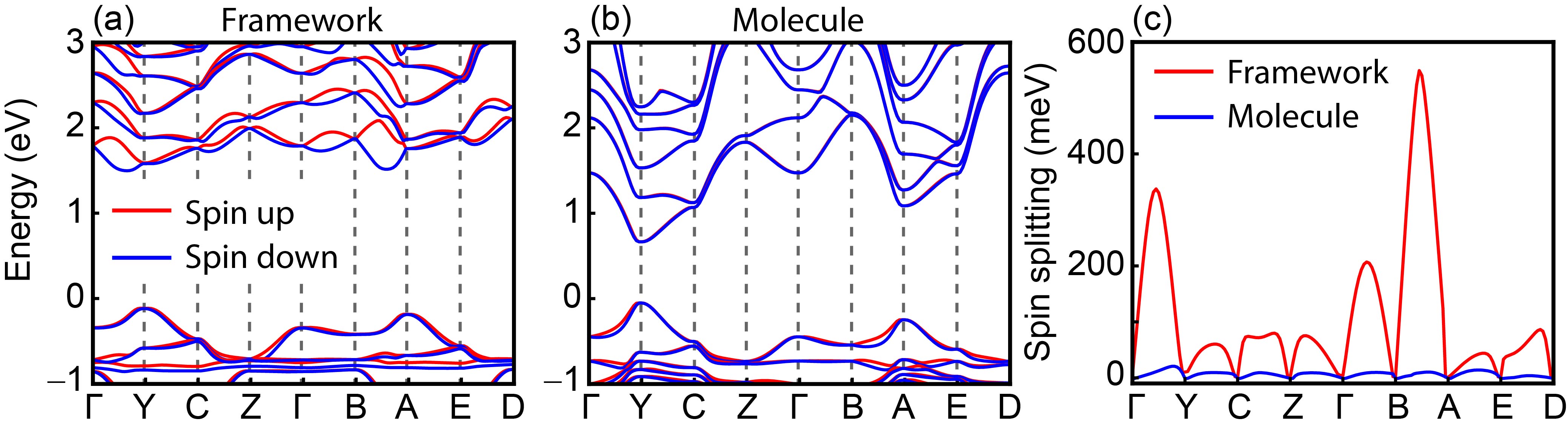}
\caption{The electronic band structure for (a) framework distortion and (b) molecule rotation of \mpb. (c) The corresponding SS of framework distortion and molecule rotation of \mpb\ for CB.}
\label{Fig8}
\end{figure}

\begin{figure}[h!]
\centering
\includegraphics[width=1.0\textwidth]{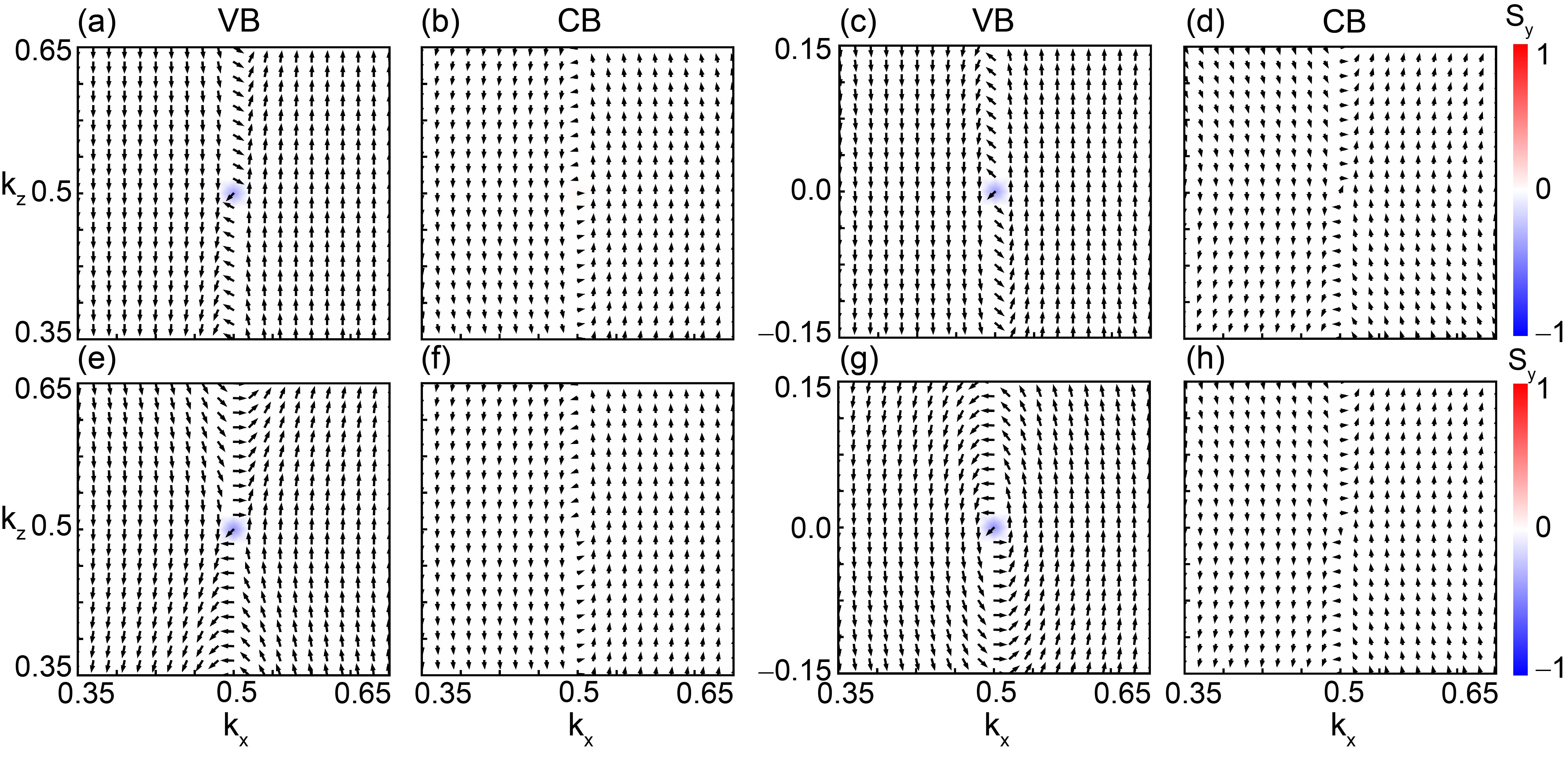}
\caption{The spin texture of VB and CB of (a)-(d) \mpb\ and (e)-(h) \mpc\ around A and Y high symmetry points in the k$_x$-k$_z$ plane, obtained from DFT.}
\label{Fig9}
\end{figure}

\begin{figure}[h!]
\centering
\includegraphics[width=1.0\textwidth]{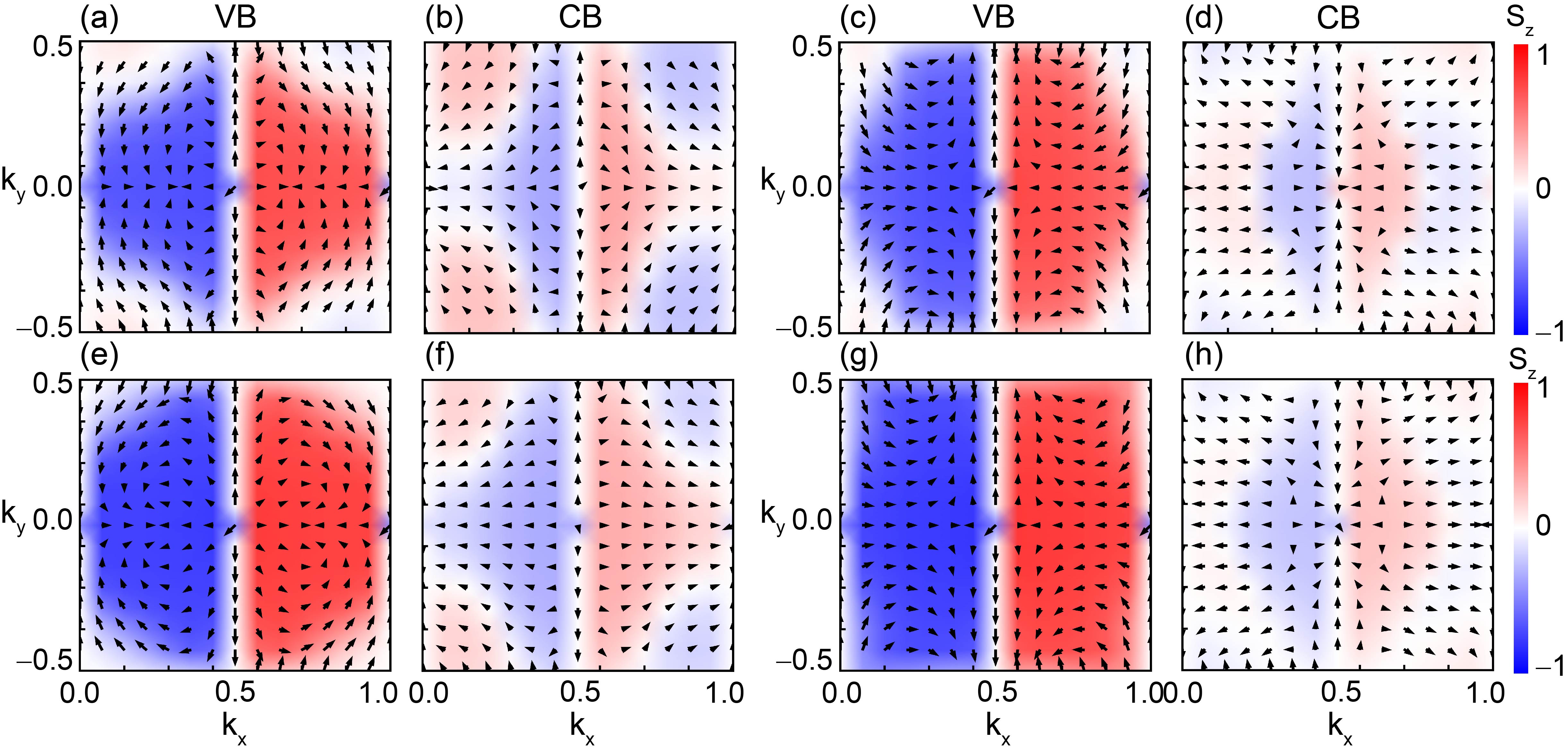}
\caption{The spin texture of VB and CB of (a)-(d) \mpb\ and (e)-(h) \mpc\ around A and Y high symmetry points in the k$_x$-k$_y$ plane, obtained from DFT.}
\label{Fig10}
\end{figure}

\begin{figure}[h]
\centering
\includegraphics[width=1.0\textwidth]{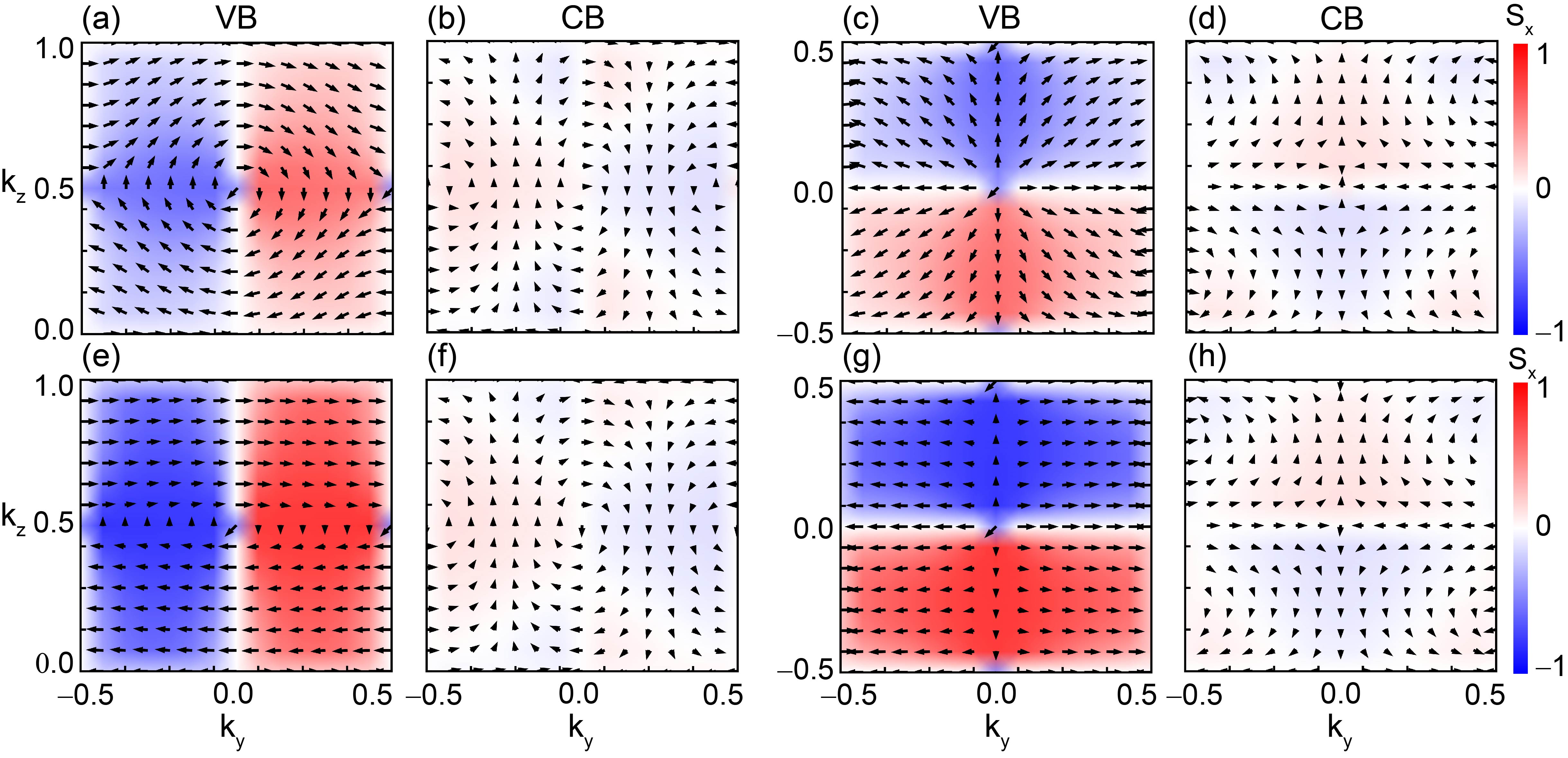}
\caption{The spin texture of VB and CB of (a)-(d) \mpb\ and (e)-(h) \mpc\ around A and Y high symmetry points in the k$_y$-k$_z$ plane, obtained from DFT.}
\label{Fig11}
\end{figure}

\begin{table}[h!]
\begin{center}
\caption{Symmetry operators and corresponding invariants for C$_2$ point group symmetry.}
\begin{tabular}{p{3cm} p{3cm} p{3cm} p{7cm}}
\hline
Symmetry & ($k_x$, $k_y$, $k_z$) & ($\sigma_x$, $\sigma_y$, $\sigma_z$) & Invariants \\
\hline
 T = i$\sigma_y$ $K$  & ($-k_x$, $-k_y$, $-k_z$)     & ($-\sigma_x$, $-\sigma_y$, $-\sigma_z$)     & $\overline{k_x\sigma_x}$, $k_x\sigma_y$, $\overline{k_x\sigma_z}$, \\

 & & & $k_y\sigma_x$, $\overline{k_y\sigma_y}$, $k_y\sigma_z$, \\
 & & & $\overline{k_z\sigma_x}$, $k_z\sigma_y$, $\overline{k_z\sigma_z}$ \\
 $C_{2y}$= $-i\sigma_y$  & ($-k_x$,  $k_y$, $-k_z$)     & ($-\sigma_x$,  $\sigma_y$, $-\sigma_z$)     & $\overline{k_x\sigma_x}$, $\overline{k_x\sigma_z}$,  \\
 & & & $\overline{k_y\sigma_y}$, \\
 & & & $\overline{k_z\sigma_x}$, $\overline{k_z\sigma_z}$ \\
 
\hline
\end{tabular}
\label{tab-s2}
\end{center}
\end{table}

\end{document}